# Cavity-enhanced radiative emission rate in a single-photon-emitting diode operating at 0.5 GHz.

David J P Ellis [1,3], Anthony J Bennett [1], Samuel J Dewhurst [1,2], Christine A Nicoll [2], David A Ritchie [2] and Andrew J Shields [1].

[1] Toshiba Research Europe Limited, 208 Cambridge Science Park, Milton Road, Cambridge, CB4 0GZ, UK

[2] Cavendish Laboratory, University of Cambridge, J. J. Thomson Avenue, Cambridge, CB3 0HE, UK.

[3] e-mail: david.ellis@crl.toshiba.co.uk

**Abstract**

We report the observation of a Purcell enhancement in the electroluminescence decay rate of a single quantum dot, embedded in a microcavity light-emitting-diode structure. Lateral confinement of the optical mode was achieved using an annulus of low-refractive-index aluminium oxide, formed by wet oxidation. The same layer acts as a current aperture, reducing the active area of the device without impeding the electrical properties of the p-i-n diode. This allowed single photon electroluminescence to be demonstrated at repetition rates up to 0.5 GHz.

The realisation of high efficiency sources of single photons or entangled pairs would benefit the fields of quantum communication and quantum computing [1]. One attractive approach uses individual semiconductor quantum dots (QDs) integrated into conventional light-emitting diodes (LEDs) [2, 3]. Such sources are compact, robust and remove the requirement of a pump laser. It has, however, proven difficult to integrate a high-Q cavity with low mode volume into an LED structure with electrical contacts. Here we present a single photon LED incorporating an oxide-confined microcavity. For the first time we see cavity enhancement of the radiative decay rate due to the Purcell effect in an electrically pumped device. We demonstrate efficient single photon emission from an individual QD under pulsed electrical injection up to 0.5 GHz.

In recent years, the generation of single photons and entangled pairs from individual semiconductor QDs has been demonstrated [1,4], both of which are essential to allow photonic information processing schemes to be fully exploited. If an InAs quantum dot is embedded in bulk GaAs, the refractive index contrast at the air/GaAs interface ($n_{GaAs} \sim 3.5$ at 900 nm) allows only a small proportion of the emitted photons to escape through the top of the crystal. One solution is to incorporate the QD into a cavity and to couple the QD emission to an optical mode [5]. In addition, the radiative decay rate is enhanced, relative to a QD in free space, by an amount $F_P$, the Purcell factor, such that

$$F_P = \frac{\gamma_C}{\gamma_0} = \frac{3}{4\pi^2}\left(\frac{\lambda_C}{n_{eff}}\right)^3\left(\frac{Q}{V}\right)$$

where $\gamma_c$ and $\gamma_0$ are the radiative decay rates of a QD resonant with a cavity mode of wavelength $\lambda_c$ or in free space, respectively. $V$, $Q$ and $n_{eff}$ are the mode volume, quality factor and effective refractive index of the structure. A high Purcell factor will lead to higher collection efficiencies and shorter state lifetimes, the latter being important for high repetition rate devices and for guaranteeing indistinguishable photons [6].

To date, all demonstrations of such a Purcell enhancement in the emission of single QDs have relied upon optically exciting carriers in the semiconductor by means of a pump laser. Such systems are expensive, can be bulky, and necessitate the inclusion of additional optical components to focus the laser beam onto the sample. In addition, the repetition rate of the emitted photons is limited by the repetition rate of the pump laser. In contrast, the electrical device discussed here eliminates the need for the pump laser and associated components and photon emission rates are limited only by the driving electronics and the (enhanced) radiative decay rate of the QD state. Unlike previous work on optically pumped, oxide-confined microcavities [7, 8], the device described here proves that it is possible to combine electrical injection with a high-Q cavity such that a Purcell effect can be realised.

In conventional small mode volume semiconductor cavities such as micropillars or photonic crystals, the Q-value is sharply reduced by sidewall roughness and coupling into leaky modes. Oxide confinement offers a solution to this problem. Wet oxidation allows the formation of rings of low refractive index ($n \sim 1.5$) dielectric within etched structures. Figure 1a shows an optical microscope photograph of a square mesa that has been etched and partially oxidised. The oxide region can be observed as a region of differing colour due to the refractive index contrast between the oxide and unoxidised AlGaAs in the centre of the mesa. It is important to note that we observe a circular profile for small apertures, even if the mesa is square.

The oxide aperture serves a dual purpose. Firstly it reduces the electrically active region of the device and hence the number of electrically active QDs in the sample [9, 10]. Secondly as a region of low refractive index which results in strong lateral confinement of the cavity's optical modes as shown in the numerical simulations of the electric field distribution (Figure 1b) [11, 9, 12]. This approach allows the LED to be physically large (tens of microns wide) and therefore easy to contact electrically, whilst containing a micron or sub-micron-scale active area and optical mode diameter. As a result, such devices can be fabricated using conventional photolithography.

The sample was grown by molecular beam epitaxy and consisted of a low density of InAs self-assembled QDs, located at the centre of a $\lambda$ thick GaAs cavity. GaAs/$Al_{0.9}Ga_{0.10}$As distributed Bragg reflectors consisting of 17 (25) mirror periods were placed above (below) the cavity /oxidation region, respectively. The upper mirror was n-doped with silicon and the lower mirror was p-doped with carbon to allow the electrical injection of carriers into the cavity region. The oxidation layer consisted of three layers of $Al_xGa_{1-x}$As with Al-fractions of 70%, 90% and 100% [10]. After oxidation, these layers resulted in a $\lambda/2$ thick layer of oxide with a linear tapered profile. Such a tapered profile was chosen in preference to an oxide region with a square edge profile in order to realise a more gradual change in radial refractive index and thus to minimise scattering losses from the aperture's edge [10, 11, 12].

Processing was carried out with standard optical photolithography techniques. Device fabrication began with a series of etches using a $SiCl_4$-based reactive ion etch process to define the mesa profile illustrated schematically in Figure 1c. Wet oxidation was carried out at 400 °C to produce a ring of oxide. The diameter of the resulting aperture was controlled by varying the oxidation time. Layers of silicon nitride, PdGe, InZn and TiAu were subsequently deposited and, where appropriate, annealed to form ohmic contacts to the n- and p-doped regions of the device, respectively.

Figure 1d shows a scanning electron microscope image of a partially completed device. The upper section (above the active region) was 8 μm wide whilst the lower part (below the active region) had a width of 18 μm. The oxidation time was controlled to produce an oxide aperture with a diameter of ~ 2.4 μm, based upon calibrations using test samples. The resulting devices exhibit a standard diode current-voltage characteristic with a well-defined turn-on at ~1.5 V. Below threshold, leakage currents of 1-10 pA were measured. Above threshold electroluminescence (EL) from the embedded QDs was observed.

Measurements were performed in a continuous-flow liquid-helium optical cryostat. Optical pumping was achieved using a mode-locked Ti:Sapphire laser whilst electrical injection was performed using a Keithley Source-Measure-Unit and an Agilent pulse generator. Emission, collected using an infinity-corrected microscope objective of numerical aperture 0.5, was dispersed by a grating spectrometer and analysed by either a liquid nitrogen-cooled charge-coupled device camera, or a pair or silicon avalanche photo-diodes.

The mode structure of the cavity may be observed in the electroluminescence spectra of the LED recorded for high injection currents, for which broadband emission is created in the cavity layer. A typical mode structure is shown in Figure 2a. Here the longest wavelength ($HE_{11}$) mode was observed at around 935 nm and typically exhibited Qs in the range 1500 – 2800. These modes are observed to be blue-shifted by ~2.45 nm relative to the planar cavity mode in a device without oxidation. Using a simple "waves-in-a-box" model, it can be shown that the wavelength of the $HE_{11}$ mode varies with radius such that $\lambda(r) = \lambda_0 - C/(n_{eff}d)^2$ where $\lambda_0$ is the wavelength of the planar cavity mode and $C$ is a constant. Using the experimentally determined blue-shift, we calculate the diameter of the oxide aperture to be 2.39 μm, which is consistent with the target value determined from the mesa radius and the rate and duration of oxidation.

Figure 2b shows a 20 K EL spectrum of a diode recorded for a low injection current of 100 nA. For low injection currents, the EL consists of a number of sharp lines due to the QD excitonic transitions and a broader line at 935 nm due to the $HE_{11}$ cavity mode. The spectral dependence of the emission suggests that two quantum dots (labelled A and B) are electrically excited under the aperture in this diode. By studying the current dependence of the intensities of the lines and their polarisation splitting, we assign the observed lines to the neutral and charged transitions of dot A and B, as indicated in Figure 2b.

By varying the temperature of the device, it was possible to tune any of the three of the emission lines onto resonance with the cavity mode as shown in Figure 2c. In each case, an enhancement in the emission intensity of the state is observed. However, of the three lines, the $X^+_B$ line showed the largest enhancement (more than a factor of 20, Figure 2d), possibly indicating that $QD_B$ has better spatial alignment to the cavity mode than $QD_A$.

The decay rate of the electroluminescence transient from $X^+_B$ state, excited by 200ps wide voltage pulses was then measured (Figure 2e). On resonance we measured a maximum EL decay rate of $(1.59 \pm 0.02) \times 10^9$ s$^{-1}$. As the electroluminescence decay rate can be modified by tunnelling of carriers from the dot due to the falling edge of the applied voltage pulse, we performed a time resolved photoluminescence measurement on the same device to accurately determine the radiative recombination rate. For these measurements the diode was dc biased just below turn-on to prevent carriers tunnelling from the dot. On resonance we measured a radiative recombination rate of ~1.4 ns$^{-1}$.

Figure 2f plots the variation in the measured radiative recombination rate as the dot is tuned through the cavity wavelength by changing the device temperature. The observed enhancement in the radiative decay rate corresponds to a Purcell factor, $F_P$, of $(2.49 \pm 0.05)$. We have numerically modelled the structure and calculated the volume of the confined mode to be 1.42 μm$^3$. When combined with the experimentally determined mode wavelength and quality factor, we estimate the value of Fp to be $(2.8 \pm 0.7)$, which is consistent the measured value.

Coupling the QD emission to the cavity mode also allows photons emitted from the device to be more efficiently coupled into a measurement system. In order to evaluate this collection efficiency, the combined detection efficiency of our system (optics, spectrometer and charge coupled device camera) for photons at this wavelength was first measured and found to be $(1.00 \pm 0.04)$ %. The device was driven close to the saturation current (12 nA, see Figures 3a, 3b). H- and V- polarised spectra were recorded and the area integrated (Figure 3c). The experimentally determined losses were then used to calculate the number of photons captured by the first lens assuming $g^{(2)}(0) = 0$ and that all decay occurs via this state. The collection efficiency into NA = 0.5 was evaluated by dividing the number of captured photons per second by the repetition rate of the source (80 MHz), yielding a value of $(14 \pm 1)$ % at the resonance temperature. This represents a three-fold improvement over previously reported planar cavity structures [13] and a 28-fold improvement over simple structures without a cavity [2,9,14,3].

In order to verify that the emission originates from a single quantum state, the photon emission statistics were studied using a Hanbury-Brown and Twiss coincidence arrangement consisting of a beam splitter, two avalanche photo-diodes and time-correlated counting electronics. Figure 3b shows the recorded auto-correlation histogram obtained at saturation current. The presence of well-resolved peaks confirms that the photon emission has been successfully regulated by the application of the voltage pulses. In addition there is a strong suppression of counts in the zero time-delay peak, confirming anti-bunched emission even at saturation. To assess the background contribution to this measurement, a spectrum was recorded and fitted with multiple Lorzentzian line shapes to determine the contribution from the QD and the cavity mode. The entirety of the residual area of the zero-delay peak is consistent with the background level suggested by the fitted line shapes shown in Figure 3c, confirming that we are obtaining close to ideal single photon emission from the $X^+$ state at saturation.

A significant advantage of such a single-photon-emitting-diode over optically pumped devices is that the emission pulse rate is no longer limited by the repetition rate of the pump laser. To demonstrate this, the diode was then excited using pulses at a rate of 0.5 GHz. The resulting auto-correlation histogram is shown in Figure 4b. The peaks are broadened by the decay rate of the $X^+$ state and the finite response time of our system and, as a result, neighbouring peaks in the histogram begin to overlap. However, we again observe a strong suppression of coincidence counts in the zero delay peak together with no increase in $g^{(2)}(0)$ when compared to 80 MHz operation at low current (Figure 4a). Even higher operating frequencies would be possible with improved cavities and faster single photon detectors.

To conclude, we have demonstrated an efficient single photon LED demonstrating a Purcell enhancement of the radiative decay rate of 2.49 ± 0.05. The collection efficiency of 14 % is the highest value reported for an electrically driven single photon source. The use of a tapered oxide aperture simultaneously provided electrical and optical confinement within the structure. Pulsed electrical injection allows single photon emission with repetition rates up to 0.5 GHz without any degradation in $g^{(2)}(0)$. By implementing higher-Q devices and by improving the QD – cavity mode coupling, larger Purcell enhancements could be achieved.

**Acknowledgements**

This work was supported by the European Commission under Integrated Project SECOQC, the Integrated Project Qubit Applications (QAP) funded by the IST directorate as Contract Number 015848 and Framework Package 6 Network of Excellence SANDiE.

Correspondence and requests for materials should be addressed to D.J.P.E.

D.J.P.E would also like to thank Mark Stevenson, Martin Ward for fruitful discussions and Raj Patel for Figure 1c.

**Figures**

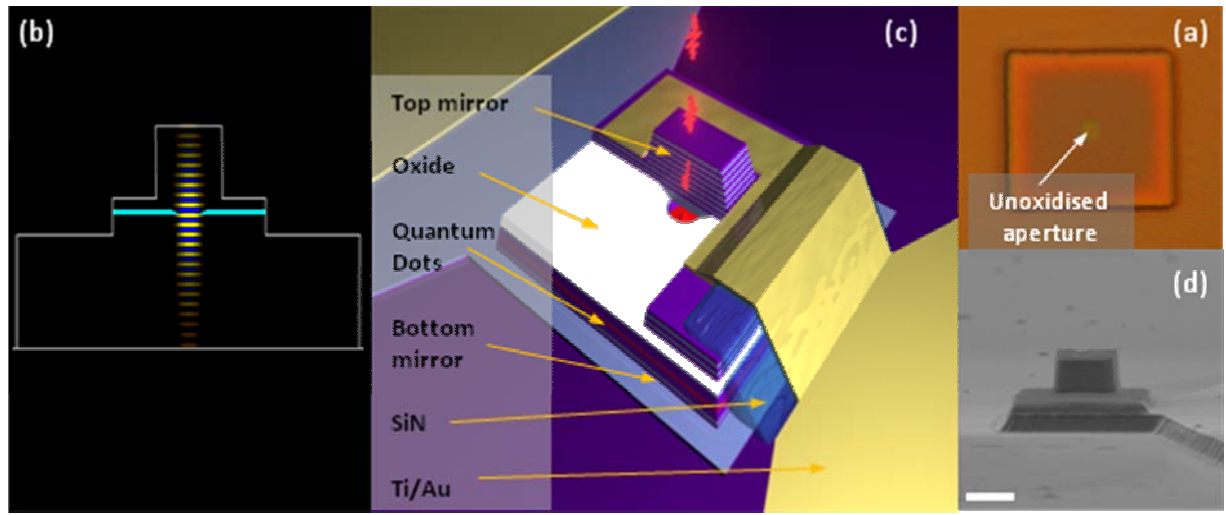

**Figure 1** Oxide-confined microcavity single-photon-emitting diode (SPED).

**(a)** Optical photograph of a partially oxidised square mesa. As the aperture closes off, the square profile becomes smoothed to form a circular aperture. The colour difference originates from the refractive index contrast between oxidised and un-oxidised regions. The oxidation rate is dependent upon AlGaAs composition. Oxidation proceeds more slowly in the 90% AlGaAs layer, compared to the core of the aperture layer, resulting in the second oxide front visible as a light coloured region around the edge of the mesa. **(b)** Numerical simulation illustrating confinement of the internal electric field in the oxide-confined structure. The white lines represent mesa edges. The thick light blue regions represent an oxide annulus. **(c)** Schematic of an oxide-confined microcavity SPED. **(d)** Scanning electron micrograph of a device after mesa etching has been completed. The scale bar represents 5 μm.

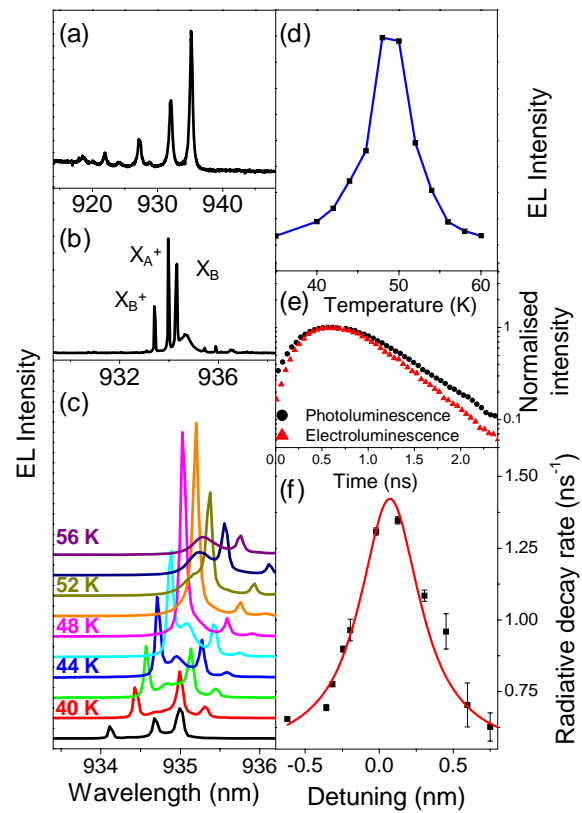

**Figure 2** Purcell-enhanced quantum dot emission.

(a) Emission spectrum at high power. Cavity modes are visible. (b) Low current spectrum recorded at 20 K. Emission lines from two quantum dots (labelled A and B) are visible in addition to the $HE_{11}$ cavity mode at slightly longer wavelength. (c, d) Temperature tuning. Quantum dot emission lines exhibit enhanced intensity when tuned onto resonance with the cavity mode. The $X^+$ emission line from quantum dot-B exhibits more than a 20-fold increase in intensity when on resonance. (e) Decay curves recorded at zero detuning under photoluminescence (black) and electroluminescence (red). (f) Enhancement in the radiative decay rate for $X^+_B$ state. When on resonance, the radiative decay rate is enhanced with a Purcell factor of 2.5.

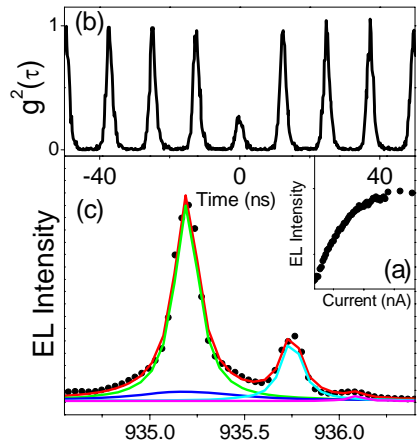

**Figure 3** Emission properties of the device.

**(a)** Electroluminescence intensity as a function of injection current when modulating the device emission with 200 ps-wide pulses at a repetition rate of 80 MHz. Intensity saturates for a time-averaged current of 12 nA. **(b)** Auto-correlation histogram recorded under these conditions. A strong suppression of the zero-delay peak is observed. Such suppression is consistent with the Lorentzian fits to a charge-coupled device camera spectrum under these conditions shown in **(c)**. The green, cyan and magenta lines fit the three quantum dot features whilst the blue line is a fit to the cavity mode.

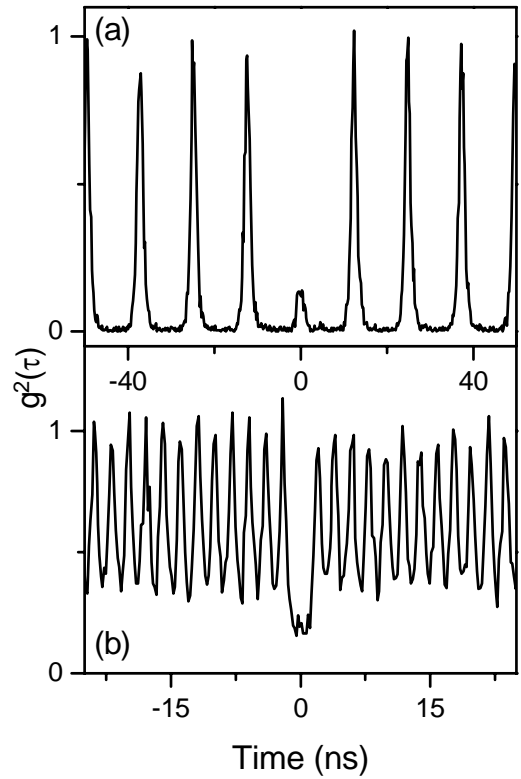

**Figure 4** Auto-correlation histograms at 48K.

**(a)** Measurement performed with the device driven at a repetition rate of 80 MHz with a time-averaged current of 1 nA. A strong suppression of the zero-time-delay peak confirms the emission of a single photons from the quantum dot with $g^{(2)}(0) = 0.17$ **(b)** Measurement repeated at a repetition rate of 0.5 GHz, demonstrating the flexibility of the electrically driven single-photon-emitting diode. Increasing the repetition rate does not increase $g^{(2)}(0)$.